\documentstyle[aps,prb,twocolumn, epsf]{revtex}
\begin{document}
\wideabs{
\title{Phase separation in La$_{0.5}$Ca$_{0.5}$MnO$_3$ doped with $1\%$ $^{119}$Sn
detected by M\"ossbauer spectroscopy}
\author{A. Simopoulos, G. Kallias, E. Devlin, M.Pissas}
\address{Institute of Materials Science, National Center for Scientific Research "Demokritos", 153 10 Athens, Greece}
\date{\today }
\maketitle

\begin{abstract}
$1\%$ $^{119}$Sn-doped La$_{0.5}$Ca$_{0.5}$MnO$_3$ was studied by M\"ossbauer spectroscopy,
magnetic moment and resistivity measurements.
The M\"ossbauer spectra below the charge-ordering temperature are explained with
ferromagnetic (FM), antiferromagnetic (AF), and ferromagnetic spin cluster (CL) components.
The magnetic and thermal hystereses of the relative intensities of the components observed
in the M\"ossbauer spectra, and of the bulk properties such as magnetic moment and electrical
resistivity, in the temperature range $125-185$ K, are characteristic of phase equilibrium in a first-order
transition, i.e. of phase separation in the system below the charge-ordering (CO) transition.
The cluster component displays a significant hyperfine field up to $\sim 125$ K. Above this temperature
it exhibits superparamagnetism, becoming the dominant component above the
charge-ordering transition.
These results are discussed in the framework of recent investigations of the
manganite system with other techniques which also show phase separation.
\end{abstract}
}

\section{Introduction}
The Lanthanum manganite series La$_{1-x}$Ca$_x$MnO$_3$ presents a remarkably
complicated phase diagram as the number of Mn$^{4+}$ changes. \cite{wollan55,schiffer95,ramirez96}
For $x\leq 0.45$ the ground state is a FM metal and above $x=0.55$ a charge ordered AF insulator.
\cite{radaelli97,radaelli99}
In the intermediate doping regime, mixtures of the two phases have been observed and close to
$x\sim 0.5$ it was shown that even a small variation of the Mn$^{4+}$ content
can greatly change the properties. \cite{roy99}

The co-existence of FM and AF phases (or phase separation) in this doping regime has been predicted
theoretically \cite{nagaev98,moreo99} and witnessed by a number of techniques.
This co-existence can be expected since the energies of the two relevant interactions
(double-exchange and superexchange) are comparable.
This energetic argument is supported
by the fact that application of an external magnetic field changes the energy balance
enhancing the FM phase through a melting transition of the charge-ordered lattice. \cite{xiao97,kallias99}
In bulk magnetization measurements, the high value of the magnetic moment observed at low temperatures
implies the existence of a ferromagnetic component (its amount varies with the sample preparation
route). \cite{schiffer95,roy99,xiao97,kallias99}
Electron diffraction studies \cite{mori98,chen96} have shown that, between $95$ and $135$ K,
La$_{0.5}$Ca$_{0.5}$MnO$_3$ is
an inhomogeneous spatial mixture of incommensurate charge-ordered and ferromagnetic charge-disordered
microdomains with a size of $200-300$  \AA (chemical inhomogeneities have been ruled out with
electron microprobe analysis with a spatial resolution of $200$  \AA).
Phase separation was detected by local NMR probes ($^{139}$La, $^{55}$Mn) at all
temperatures below the first formation of the CO state in La$_{0.5}$Ca$_{0.5}$MnO$_3$.
\cite{allodi98,dho99}
M\"ossbauer specrtroscopy in $^{57}$Fe-doped  La$_{0.5}$Ca$_{0.5}$MnO$_3$ has also shown
coexistence of the two phases and manifested clearly the thermal and magnetic history dependence of their
relative fractions. \cite{kallias99}
Finally, coexistence of charge ordered AF and charge disordered FM phases
was observed below the charge-ordering transition by neutron diffraction measurements in
the $x=0.5$ \cite{kallias00} and $x=0.47$ compounds. \cite{rhyne99}

In this paper, we report a detailed M\"ossbauer study of La$_{0.5}$Ca$_{0.5}$MnO$_3$ doped
with $1\%$ $^{119}$Sn. The aim is to follow microscopically the evolution of the phases that appear
as the temperature is varied.
The latter substitutes for Mn$^{4+}$ and serves as a probe of the magnetic state of the host
lattice through the transferred hyperfine interactions from its neighbor Mn magnetic ions
in a similar fashion with the $^{139}$La probe in NMR investigations.
The results show phase separation at all temperatures below the CO temperature
as witnessed by the FM and AF components in the M\"ossbauer spectra. A third component appearing
in the M\"ossbauer spectra even at $4.2$ K is attributed to spin clusters. The relative
intensities of these components show thermal and magnetic hysteresis,indicating that the
FM-AF transition is of first-order. Hysteretic behavior is also displayed in bulk magnetic
moment and resistivity measurements.

\section{Experimental}
We prepared a  sample with nominal composition La$_{0.5}$Ca$_{0.5}$Mn$_{0.99}$Sn$_{0.01}$O$_3$
by thoroughly mixing high purity stoichiometric amounts of La$_2$O$_3$, CaCO$_3$,
MnO$_2$ and SnO$_2$ ($90\%$ enriched in $^{119}$Sn). The mixed powders were pelletized and
annealed in air at $1390^{\rm o}$ C and $1410^{\rm o}$ C for approximately $300$ h with intermediate
grindings and reformation into pellets each time. Finally, the sample was slowly cooled in
the furnace.

X-ray powder diffraction (XRD) data were collected with a D500 SIEMENS
diffractometer, using CuK$\alpha $ radiation and a graphite crystal
monochromator, from 4$^{\text{o}}$ to 120$^{\text{o}}$ in steps of 0.03$^{%
\text{o}}$ in $2\Theta $. The power conditions were set at 40KV/35mA. The
aperture slit as well as the soller slit were set at $1^{\text{o}}$.
The refinement of the x-ray diffraction patterns was carried out by the BBWS-9006
Rietveld program \cite{wiles81} using the space group Pnma.
The shape of the peaks was assumed to be Pearson VII and the background was refined
together with the structure. 

\begin{figure}[tbp] \centering
\epsfxsize=8.6cm \epsfclipon \epsffile{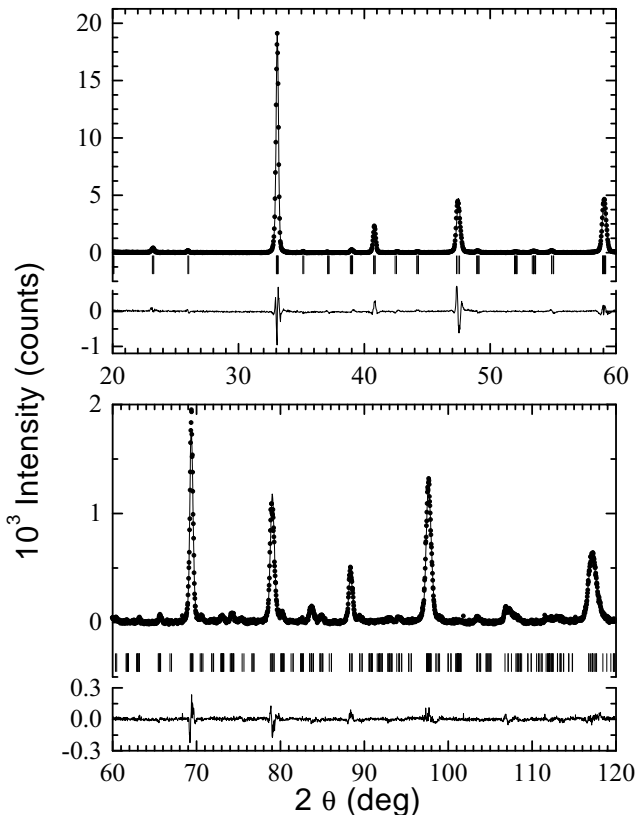}
\caption{
Rietveld refinement patterns for La$_{0.5}$Ca$_{0.5}$Mn$_{0.99}$Sn$_{0.01}$O$_3$ using x-ray powder 
diffraction data. The observed intensities are shown by dots and the calculated ones by the solid line. 
The positions of the Bragg reflections are shown by the small vertical lines below the pattern. The line at 
the bottom indicates the intensity difference between the experimental and the refined patterns.
\label{fig1} }
\end{figure}

\begin{table}
\caption{
Fractional atomic coordinates, isotropic temperature factors and occupancy factors for
La$_{0.5}$Ca$_{0.5}$\-Mn$_{0.99}$Sn$_{0.01}$O$_3$ compound using powder x-ray diffraction data.
Rietveld refinements were done in the space Group $Pnma$ with lattice parameters $a=5.4175(2)$\AA, 
$b=7.6401(2)$\AA\ and $c=5.4276(3)$\AA.
The reliability factors were $R_{p}=4.99\%$, $R_{wp}= 9.88\%$ and $R_{B}=4.06\%$.
Numbers in parentheses are statistical errors of the last significant digit.}
\label{table1}
\begin{tabular}{ccccccc}
Atom &Wyckoff & $x$ &$y$ &$z$ & $B$ & $N$  \\
     &notation&     &    &    &     &      \\
\tableline
 La& 4a &  0.02(1)&  1/4      &  0.0031(1)&  .22(3)&  0.5\\
 Ca& 4a &  0.02(1)&  1/4      &  0.0031(1)&  .22(3)&  0.5\\
 Mn& 4b &  0      &   0       &  1/2      &  .22(3)&  .495\\
 Sn& 4b &  0      &   0       &  1/2      &  .22(3)&  .005\\
 O1& 4a &  .496(1)&  1/4      &  .0450(1) &  .5 &  1.0\\
 O2&    & -.2752(1)& -.0375(1)&  0.275(2)&  .5 & 1.0\\
\end{tabular}
\end{table}

The isotropic temperature factors for the oxygen atoms were
kept constant at the same value. 
The refined pattern shown in Fig. \ref{fig1} and
the results of the refinement given in Table \ref{table1} show the high crystalline quality of
our sample.
The lattice parameters obtained $a=5.4175(2)$ \AA, $b=7.6401(2)$ \AA\  and $c=5.4276(3)$ \AA\ are very 
close to those of the undoped $x=0.5$ compound.

DC magnetization measurements were performed in a SQUID magnetometer (Quantum Design) for
fields up to $50$ kOe. ac susceptibility measurements were performed at zero dc field by
means of a laboratory-constructed probe at a frequency of $317$ Hz and with an ac external field of
amplitude $h_{ac}=1.5$ Oe. Four-probe resistivity measurements with and without a magnetic
field were performed on a sintered bar in the temperature range $4.2$-$320$ K.
M\"ossbauer spectra were recorded with a conventional constant acceleration spectrometer
with a $5$ mCi CaSnO$_3$ ($^{119}$Sn) source moving at room temperature (RT), while the absorber was kept
fixed in a variable temperature cryostat equipped with a superconducting magnet ($65$ kOe)
in a geometry perpendicular to the $\gamma$-ray beam. The calibration of the spectrometer
was made with $\alpha$-Fe and $\alpha$-Fe$_2$O$_3$ absorbers and isomer shifts are quoted relative to
$\alpha$-Fe at room temperature.

\section{Results and Discussion}
\subsection{Magnetic moment and resistivity measurements}
\begin{figure}[tbp] \centering
\epsfxsize=8.6cm \epsfclipon \epsffile{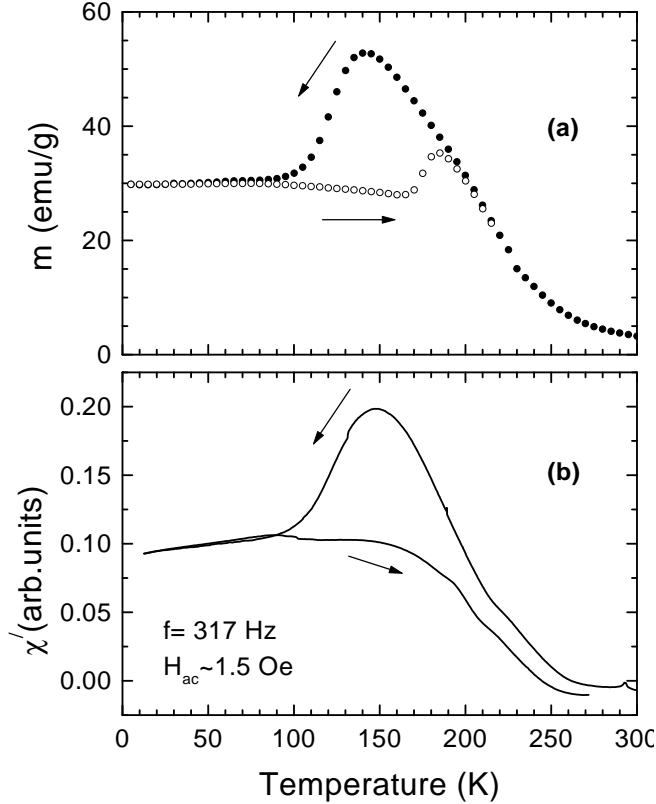}
\caption{(a) The dc magnetic moment upon field-cooling and field-cooling warming in $H=10$ kOe and
(b) the real part of the ac-susceptibility in zero dc magnetic field for
La$_{0.5}$Ca$_{0.5}$Mn$_{0.99}$Sn$_{0.01}$O$_{3}$.
\label{fig2}}
\end{figure}

Figure \ref{fig2}a shows bulk magnetic moment $m$ vs $T$ measurements in field-cooling (FC) and
field-cooling warming (FCW) modes in an external magnetic field of $10$ kOe. The sample
undergoes a rather broad paramagnet-to-ferromagnet transition at $\sim 245$ K (defined
in a measurement at $50$ Oe).

On cooling, $m$ increases smoothly with a maximum magnetic moment around $140$ K. Below
this temperature the bulk magnetic moment drops to a value that is approximately
$55$\% of the maximum value and then remains constant down to $4.2$ K.
The value of $m$ at $4.2$ K is rather large and certainly cannot be explained by an
antiferromagnetic or canted antiferromagnetic phase. The presence of a ferromagnetic
phase is needed.
Warming from $4.2$ K shows a strong hysteresis in the region
$100$-$185$ K, indicating the first-order nature of this transition.
Under an applied field of $5$ T the transition is smeared out and the bulk magnetic moment
reaches $80$\% of its full FM saturation value at low temperatures
(for an ideal ferromagnetic state in the sample it is $101.5$ emu/g or $3.5$ $\mu _B$).
The ac susceptibility measurement shown in fig. \ref{fig2}b is in agreement with the dc data.

\begin{figure}[tbp] \centering
\epsfxsize=8.6cm \epsfclipon \epsffile{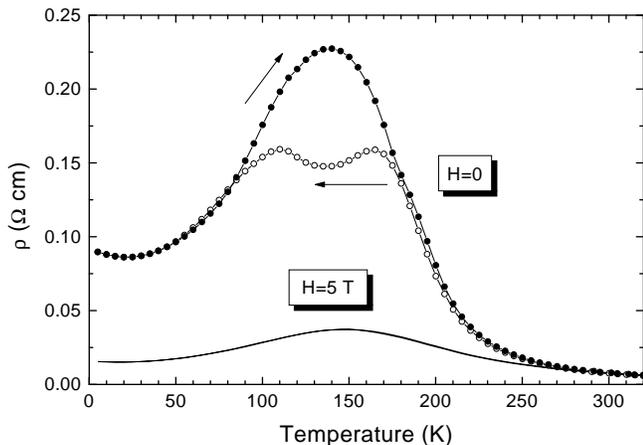}
\caption{ The temperature dependence of resistivity at $H=0$ (open and filled circles) and
$H=50$ kOe (solid line).
\label{fig3} }
\end{figure}

Resistivity measurements in zero and $5$ T fields performed following the same sequence as
the magnetic measurements are shown in fig. \ref{fig3}.
At zero field, on cooling, the resistivity increases below $250$ K and shows two small
maxima at $T\sim 160$ and $\sim 110$ K, whereas on warming a broad peak appears 
at $T \sim 140$ K, showing in strong hysteresis. In a $5$ T measurement the hysteresis 
disappears and only a broad maximum centered around $150$ K is observed.
The magnetoresistance effect $(\rho(0)-\rho(5 T))/\rho(0)$ is $\approx 85\%$ at the maximum
of the resistivity curve (around 140 K).

The maximum value of the resistivity observed at $T\sim 140$ K indicates that above this 
temperature "activated" conduction occurs and below this temperature conduction takes place 
probably through a percolation path created by FM regions in the sample, in agreement with 
the magnetization data

In a recent study Roy et al. \cite{roy99} have investigated the region around $x=0.5$ and
found differences between the $x$ value and the Mn$^{4+}$ fraction which reflect on the
magnetization and resistivity behavior of the corresponding samples. Comparing our bulk
measurements with their detailed data we can conclude that the Sn-doped sample is
slightly below $x\sim 0.5$ (e.g. $x=0.49$) in the magnetic ($T$ vs $x$) phase diagram of the
La$_{1-x}$Ca$_x$MnO$_3$ manganites.

\subsection{ M\"ossbauer measurements}
\begin{figure}[tbp] \centering
\epsfxsize=8.6cm \epsfclipon \epsffile{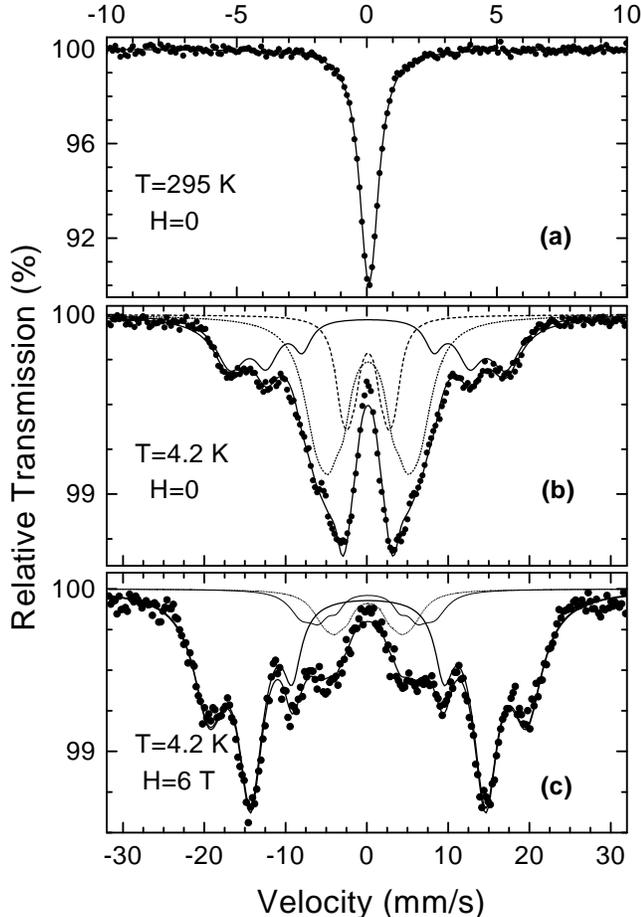}
\caption{M\"ossbauer spectra (a) at room temperature, (b) at $T=4.2$ K in zero applied
field, and (c) at $T=4.2$ K after field-cooling in an applied field of $60$ kOe.
\label{fig4}}
\end{figure}

Figure \ref{fig4} shows M\"ossbauer spectra at 300 K, $4.2$ K and $4.2$ K after field-cooling
in the presence of an external field of $60$ kOe. The results of the fitting at 4.2 K
are presented in Table \ref{table2}. The RT spectrum consists of a single absorption
line with linewidth $\Gamma=0.78$ mm/s and isomer shift $\delta=0.10$ mm/s.
This latter value corresponds to a $4+$ valence state of Sn, indicating that Sn ions
substitute for Mn$^{4+}$. The $4.2$ K spectrum is fitted with three magnetic components,
one with a large hyperfine field ($H_{hf}=249$ kOe) and the other two with smaller
hyperfine fields ($93$ and $47$ kOe).
The FC spectrum shows a dramatic increase of the high field component at the expense of the low field
components. An analogous increase (actually a reversal of the FM and AF fractions) was observed in
Fe-doped La$_{0.5}$Ca$_{0.5}$MnO$_3$, \cite{kallias99} and was attributed to the "melting" of the
CO-AF state to a FM state.

The fact that Sn nucleus senses transferred hyperfine field from the overlapping of the
$3$d orbitals of the Mn neighbors in the surrounding Mn octahedron with its $5$s orbitals,
leads us to the following assignment of the three components in the $4.2$ K Sn spectrum. The large field
component ($H_{hf}=249$ kOe) arises from a ferromagnetic environment where all the Mn moments are
parallel and add up to a large transferred hyperfine field at the Sn nucleus ({\it FM component}).
The $93$ kOe component arises from an antiferromagnetic environment where the antiparallel
moments contribute transferred hyperfine fields with opposite sign and which tend to cancel.
({\it AF component}). Since in the CE structure \cite{wollan55,radaelli97} there are four Mn$^{3+}$
ions with antiparallel moments and two Mn$^{4+}$ ions with parallel moments in the Mn
octahedron surrounding the Sn$^{4+}$ probe, the cancellation is not total and a hyperfine field value
of $\approx 1/3$ of the field of the FM component would be expected for the AF component.
As mentioned above, this component transforms to the FM component under the influence of cooling
in a $6$ T field. We will discuss the nature of the third component (CL) with a hyperfine
field of $43$ kOe at a later stage.

The FC spectrum at $4.2$ K (Fig. \ref{fig4}c) was modelled with one component ($H_{hf}=288$ kOe) with
line intensity ratio $3:4:1$ and two components ($H_{hf}=120 $ and $80$ kOe) modulated by a distribution of
hyperfine fields ($\Delta H=21$ kOe) with line intensity ratio $3:2:1$ (see Table \ref{table1}).

\begin{table}[tbp] \centering
\caption{Experimental values of the isomer shift $\delta $ relative to $\alpha$-Fe at RT in
mm/s, the hyperfine magnetic field $H$ in kG, the hyperfine magnetic field spread
$\Delta H$ modulating the linewidths, and the relative area of the components which appear
in the M\"ossbauer spectra as obtained from least squares fits of the M\"ossbauer spectra
of La$_{0.5}$Ca$_{0.5}$Mn$_{0.99}$Sn$_{0.01}$O$_3$ at $4.2$ K in zero field and at $4.2$ K
after FC in a magnetic field of $60$ kOe.
} \label{table2}
\begin{tabular}{ccccccccccccc}
& \multicolumn{4}{c}{FM} & \multicolumn{4}{c}{AF} & \multicolumn{4}{c}{CL} \\
& $\delta $ & $H$ & $\Delta H$ & A(\%) & $\delta $ & $H$ & $\Delta H$ & A(\%) & $\delta $ & $H$ & $\Delta H$ & A(\%) \\ \hline
ZFC & 0.15 & 249 & 28 & 30(1) &  0.15 & 93 & 24 & 50(1) & 0.15 & 47 & 10 & 20(1) \\
FC  & 0.15 & 288 & 28  & 80(3) & 0.15 & 120(5) & 21 & 10(4) &    &    &    &  \\
    &      &     &     &    & 0.15 &  80(3) & 21 &  10(4) &    &    &    &  \\
\end{tabular}
\end{table}

The relative intensity of the high field component (FM phase) increases
considerably (80\%) at the expense of the low-field components. The AF component present in the zero field
spectrum has a reduced intensity of 20\%, and the CL component is absorbed in the FM component due to the
applied field.
The two AF components describe the modulation of the net hyperfine field seen by the nucleus in the case
of a polycrystalline antiferromagnet with large magnetic anisotropy in an external field.
The net field at the nucleus varies from $H_0-H$ to $H_0+H$
($H_0$ is the internal and $H$ is the external field), causing
the M\"ossbauer spectra to be broadened. \cite{wertheim70}
The overall effect of the applied field to the system (and thus to the spectrum) is the same as that
observed in Fe-doped La$_{0.5}$Ca$_{0.5}$MnO$_3$. \cite{kallias99}

An important difference should be noted here for the hyperfine fields probed by the
$^{119}$Sn and $^{57}$Fe nuclei. The $^{119}$Sn ion is diamagnetic and detects
the vector sum of the hyperfine fields transferred from neighboring moments.
Thus opposite moments have a cancelling effect. However, for $^{57}$Fe the
hyperfine field arises from the iron ion's own moment. The polarisation of the
iron moment arises from the (super)exchange interactions of the Mn and
Fe spins. The superexchange interaction may result in ferromagnetic or antiferromagnetic alignment, but
the iron moment remains the same. Thus, parallel neigbouring spins are equivalent to
antiparallel spins, i.e. FM and AF environments result in the same hyperfine field.

\begin{figure}[b] \centering
\epsfxsize=8.0cm \epsfclipon \epsffile{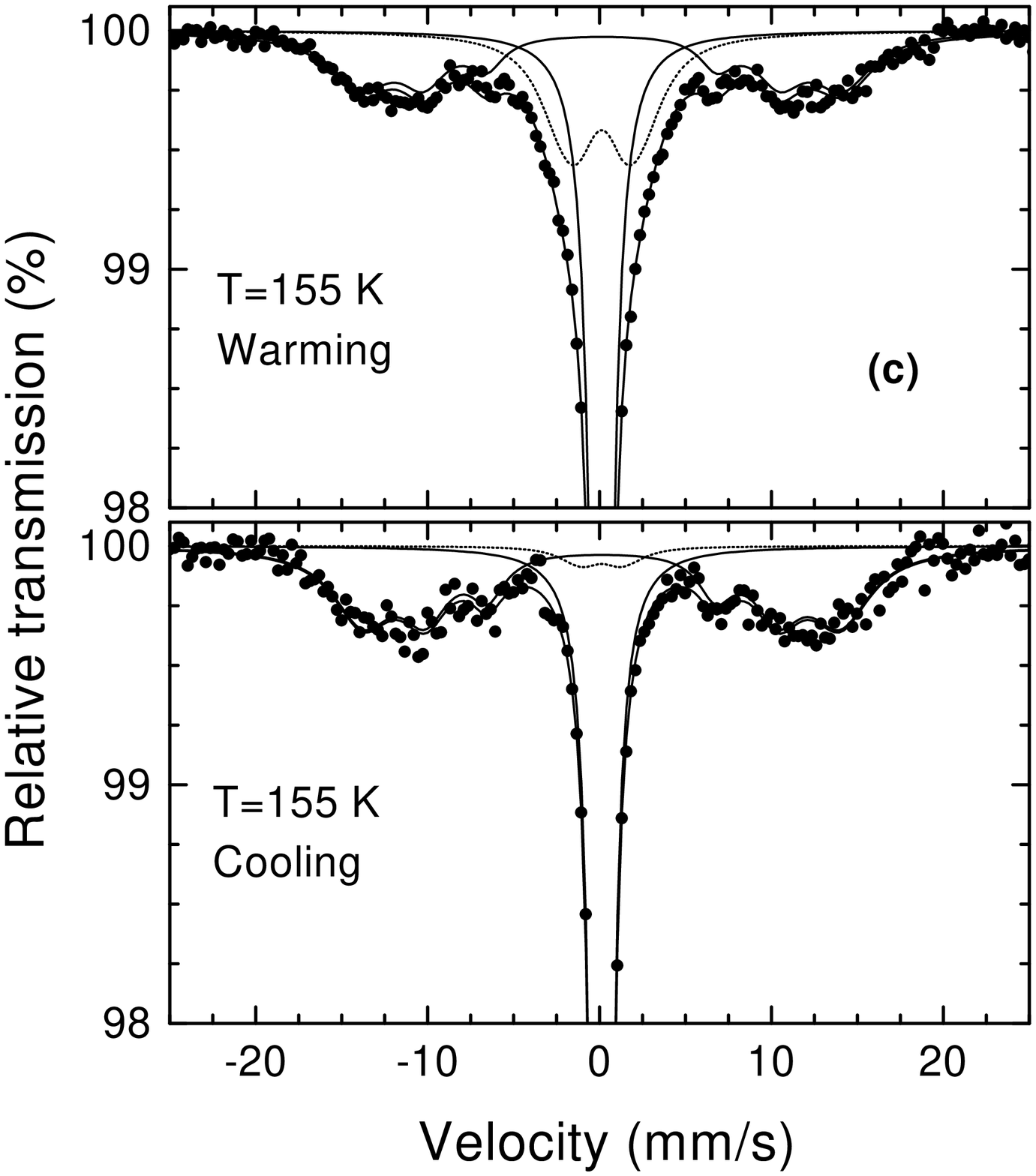}
\caption{
Representative M\"ossbauer spectra of La$_{0.5}$Ca$_{0.5}$Mn$_{0.99}$Sn$_{0.01}$O$_3$
in the temperature range $92$ to $184$ K upon (a) cooling and (b) warming.
(c) Magnified view of the spectra at $155$ K both upon warming and cooling.
\label{fig5}}
\end{figure}

We turn now to the temperature dependence of these three components.
Figure \ref{fig5} (upper panel) shows some characteristic spectra upon warming (fig. \ref{fig5}a)
and cooling (fig. \ref{fig5}b) in zero field.
The general features of the spectra do not change up to $\sim 120$ K. Above this temperature
a single peak emerges from the central region of the spectra whose intensity increases quickly as the
temperature is raised. The FM component persists up to $\sim 230$ K and above this
temperature the spectra consist of a single line only. Hysteresis is clearly evident in the central part of
these spectra. The wings which are present in the warming mode in the
temperature range $\sim 120$ to $\sim 180$ K are reduced considerably at the corresponding temperature
in the cooling mode (compare the spectra at $155$ K in the two modes in the lower panel of fig. \ref{fig5}).

We have fitted the spectra throughout the whole temperature range with the three components
of the $4.2$ K spectrum. In our fitting procedure the linewidths of the three components
were kept constant to the RT value, the isomer shift values were the same for the three
components allowing a small temperature variation due to the second order Doppler shift, and
the quadrupole interaction was kept equal to zero due to the cubic symmetry of the
Sn site. The free parameters were the hyperfine field, the intensity of each spectral
component, and a distribution of the hyperfine fields $\Delta H$. The latter arises from
small deviations of the Sn positions to which the transferred hyperfine field at the Sn
nuclei is very sensitive.

\epsfxsize=7.5cm \epsfclipon \epsffile{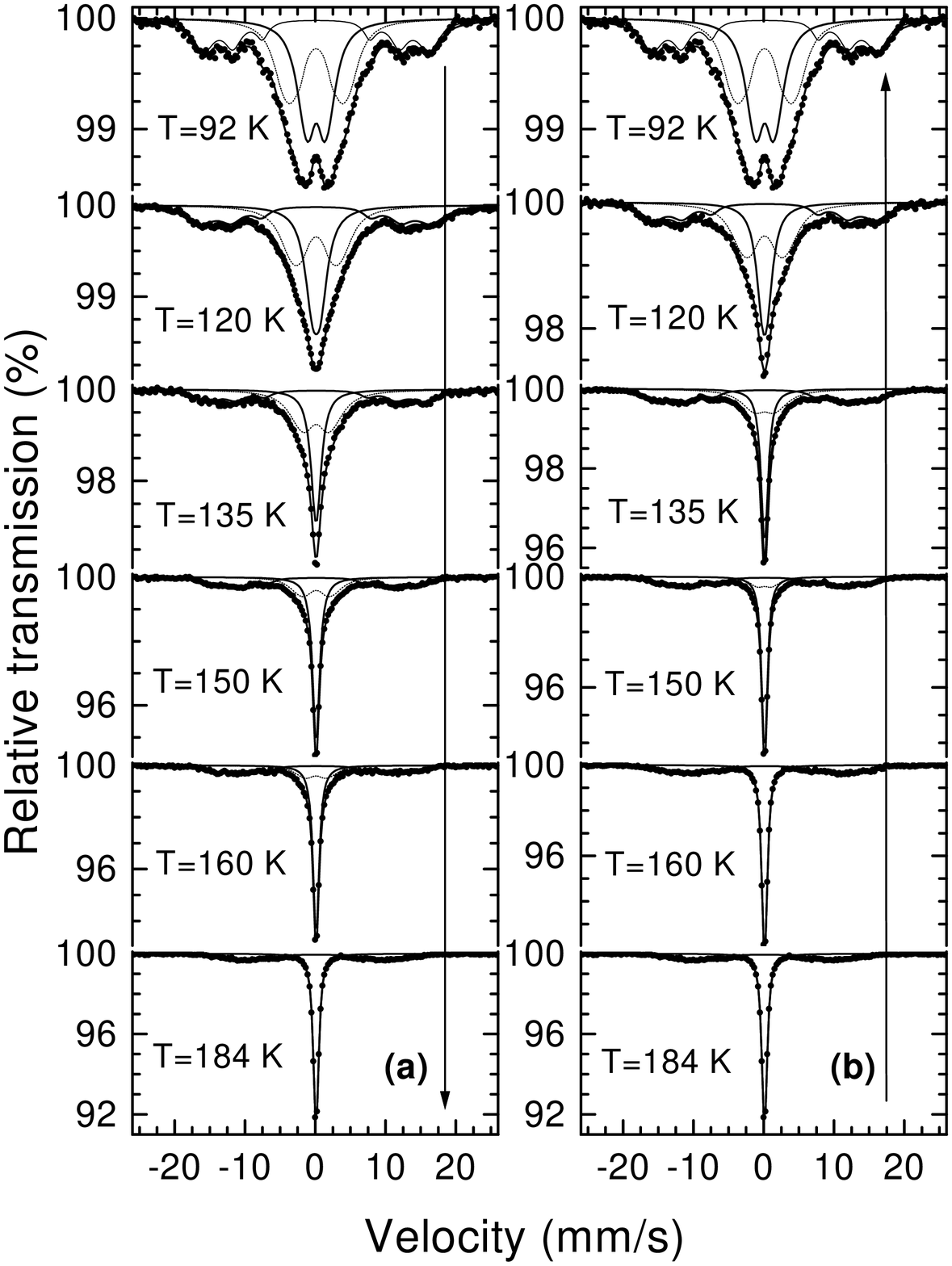}

\begin{figure}[tbp] \centering
\epsfxsize=8.6cm \epsfclipon \epsffile{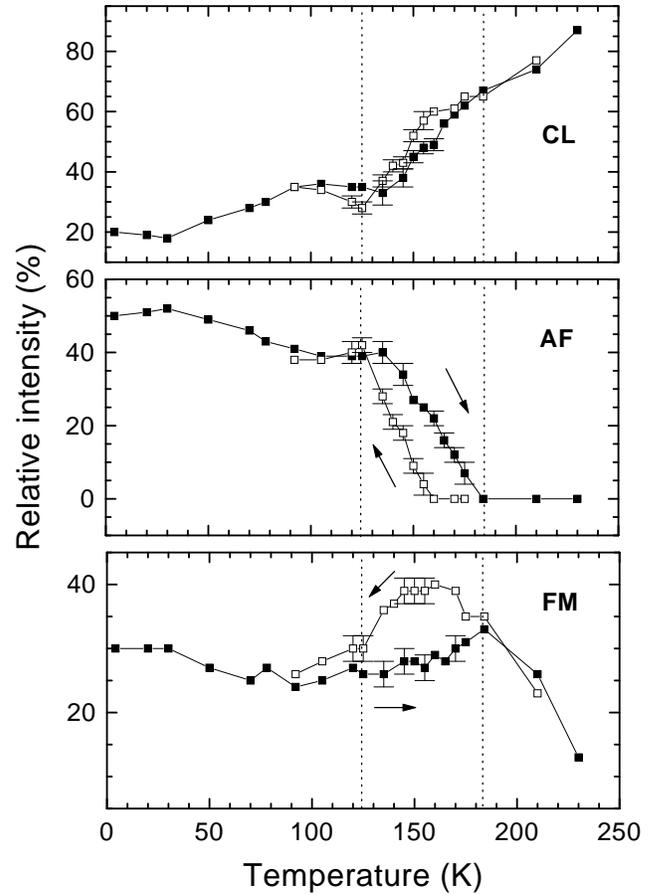}
\caption{
The temperature variation of the relative intensities of the components
which appear in the M\"ossbauer spectra of La$_{0.5}$Ca$_{0.5}$Mn$_{0.99}$Sn$_{0.01}$O$_3$.
The dashed lines mark the temperature range where hysteresis is observed.
\label{fig6}}
\end{figure}

Figure \ref{fig6} shows the main result of the paper: the temperature variation of the relative
intensity of each component.
The intensities of the FM and AF phases show hysteresis in the temperature range $120-185$ K,
as observed in the magnetic moment and resistivity
measurements described in the previous section. In particular, on warming, the FM spectral
area remains constant up to $\sim 185$ K and then drops, while on cooling it increases to
$40\%$ at $\sim 150$ K, then drops to $\sim 25\%$ at $100$ K, and remains constant down
to $4.2$ K, in agreement with the magnetization data (Fig \ref{fig2}a),
and with the La NMR results of Allodi et al. \cite{allodi98}
obtained in the cooling mode. The opposite behaviour is observed for the AF spectral area,
which disappears at $\sim 180$ K on warming and reappears at $\sim 150$ K on cooling,
thereby marking the charge-ordering temperature.
It should be noted that this phase is absent in the zero field La NMR data due to the
complete cancellation of the transferred hyperfine interactions from the neighbor Mn ions.
\cite{allodi98} Also, it cannot  be discerned in the $^{57}$Fe M\"ossbauer spectra
\cite{kallias99} since, as mentioned above, it coincides with the FM component. Thus the
$^{119}$Sn probe is ideal to detect the AF phase.

\begin{figure}[tbp] \centering
\epsfxsize=8.6cm \epsfclipon \epsffile{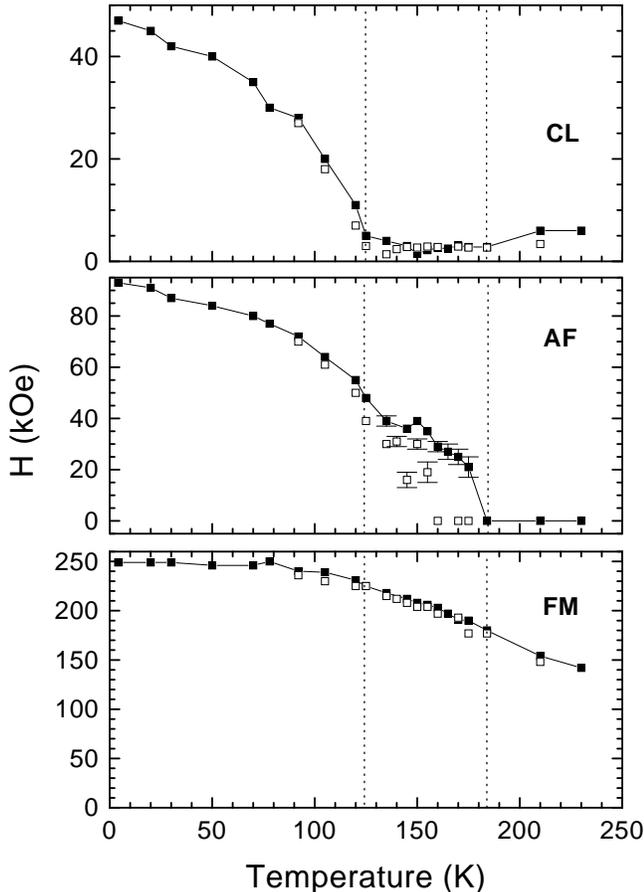}
\caption{
The temperature variation of the hyperfine fields of the components
which appear in the M\"ossbauer spectra of La$_{0.5}$Ca$_{0.5}$Mn$_{0.99}$Sn$_{0.01}$O$_3$.
The dashed lines mark the temperature range where hysteresis is observed.
\label{fig7}}
\end{figure}

Hysteresis is also observed in the temperature variation of the hyperfine field of
the AF phase (fig. \ref{fig7}). The temperature variation of H$_{hf}$ for the FM phase
follows a typical meanfield-like curve, as for the case of a ferromagnet, up to $230$ K,
above which this component disappears.
It is interesting to note that at this temperature the $H_{hf}$ value is still considerable
($142$ kOe). This discontinuity has also been observed in La NMR
experiments and it has been attributed to a first-order transition. \cite{allodi98,papavassiliou98}
This zeroing of the volume of the FM phase and not of the hyperfine field (which is contrary
to ordinary ferromagnets) implies the existence of a mixed state of FM and paramagnetic phases
near $T_c$ , thus leading to the possibility of ferromagnetic clusters. \cite{dho99}

Examining the temperature behavior of the third component (CL component) we notice
that its hyperfine field decreases smoothly with temperature and drops to zero 
at $\sim 125$K (fig. \ref{fig7}). Above this temperature
this component appears as a single paramagnetic peak with some broadening, with its intensity
increasing with temperature (see fig. \ref{fig5} and the M\"ossbauer spectra in fig. \ref{fig4}).
The zeroing of the hyperfine field of the CL component coincides with the temperature at which the 
charge-ordering is completed on cooling and is beggining to breakdown on warming.
The small saturation field ($47$ kOe) of the component does not allow its assignment to any AF spin
configuration of the Mn nearest neighbors.
Thus, we ascribe this component to ferromagnetic clusters which exhibit a reduced
hyperfine field either due to relaxation phenomena or to a reduced moment within the clusters.
These clusters are present at $4.2$ K and their spectral area remains practically constant up to $\sim 125$ K.
Above this temperature they exhibit superparamagnetic behaviour and their intensity increases rapidly at 
the expense of the AF and FM phases.
The existence of ferromagnetic clusters in manganites has been used by Moreo 
et al.\cite{moreo99b} to account for the density of states in the framework of one- and two-orbital models.
Recently, Allodi et al. \cite{allodi00} have shown by $^{55}$Mn NMR that the field-induced
FM phase in Pr$_{0.5}$Sr$_{0.5}$MnO$_3$ develops by the nucleation of microscopic ferromagnetic
domains, which may be similar in nature to the ferromagnetic clusters proposed here.

It should be noted that a paramagnetic-like component appears for $T<T_c$ in all the reported M\"ossbauer 
experiments of La manganites doped with Fe 
\cite{kallias99,pissas97,simopoulos99,tkachuk98,ogale98,simopoulos00},
Co \cite{chechersky99a,chechersky99b} or Sn, \cite{simopoulos98}
for values of Ca(Sr) doping between $0.17$ and $0.60$, which covers the phase diagram range
where the ground state is FM ($x\leq 0.45$) and AF ($x\geq 0.55$).
In addition, neutron scattering experiments for $x=0.33$, \cite{lynn96} and $x=0.17$, \cite{vasiliu97}
have detected  a quasielastic component for $T<T_c$ which becomes dominant as the temperature
approaches $T_c$. We speculate that this component is associated with ferromagnetic clusters in all these 
cases. It appears that both M\"ossbauer and neutron scattering techniques have the sensitivity to directly 
detect the formation of ferromagnetic clusters. The coincidence of the zeroing of the 
hyperfine field of the CL component with the completion and breakdown (on cooling and warming respectively) 
of the charge ordering indicates that the cluster development and growth is directly associated with charge 
ordering or charge localization.

In summarizing the M\"ossbauer data, we notice that the hysteretic range of the system is characterized by 
two temperatures, $T_1= 120$ K and $T_2= 185$ K, in agreement with the bulk measurements. In the temperature 
interval $4.2<T<120$ K the system comprises a FM phase ($\sim 30$\%), an AF phase ($\sim 50$\%) and FM 
clusters ($\sim 20$\%). The FM and AF phases display hysteresis in the range $120<T<185$ K.
The disintegration (formation) of the AF phase on warming (cooling) is clearly manifested through
M\"ossbauer spectroscopy in this temperature interval. For $T>125$ K, ferromagnetic clusters with 
superparamagnetic behavior grow rapidly in number and dominate up to the ferromagnetic to paramagnetic 
transition at $T_c=225$ K.
This picture for the La$_{0.5}$Ca$_{0.5}$MnO$_3$ system is in agreement with the phase separation models 
\cite{nagaev98,moreo99} and supplements the electron diffraction investigations \cite{mori98,chen96} and the 
La NMR data. \cite{allodi98,dho99}

\section{Conclusions}

$^{119}$Sn M\"ossbauer spectroscopy reveals that the Sn probe monitors the phase evolution during
the charge-ordering transition in La$_{0.5}$Ca$_{0.5}$MnO$_3$. Complementary bulk magnetic moment and 
resistivity measurements are in agreement with the M\"ossbauer findings. Below the charge-ordering transition, 
phase separation occurs into FM and AF phases down to $4.2$ K. The hysteresis observed shows that the FM-AF 
transition is of first order. The phase separation is associated with the existence of FM clusters which 
dominate in the system above the CO transition.


\section{Acknowledgements}
We would like to thank the M\"ossbauer group of the University of Ioannina, Greece
(Drs. A. Moukarika, Th. Bakas and V. Papaefthymiou) for lending us their new Sn source
and for preliminary meaurements, Dr. Elias Moraitakis for the ac susceptibility
measurements and Dr. G. Papavassiliou for very stimulating discussions.
Partial support by the European INTAS (Grant No. 30253/97) is also acknowledged.

\end{document}